\documentclass[letter, structabstract]{aa} % for the letters 
\usepackage{graphicx}
\usepackage{txfonts}
\usepackage{natbib}
\bibpunct{(}{)}{;}{a}{}{,}
\begin{document}
   \title{Probing the evolution of molecular cloud structure:}

  \subtitle{From quiescence to birth}

   \author{J. Kainulainen
          \inst{1}
          \and
          H. Beuther 
          \inst{1}
          \and
          T. Henning
          \inst{1}
          \and
          R. Plume
          \inst{2} 
          }

   \institute{Max-Planck-Institute for Astronomy, K\"onigstuhl 17, 69117
     Heidelberg, Germany \\
              \email{[jtkainul, beuther, henning]@mpia-hd.mpg.de}
      \and
              Department of Physics and Astronomy, University of Calgary, 2500
  University Drive NW, Calgary, Alberta T2N 1N4, Canada \\         
              \email{plume@ras.ucalgary.ca}
 }
   \date{Received ; accepted }

% \abstract{}{}{}{}{} 
% 5 {} token are mandatory
 
  \abstract
  % context heading (optional)
  % {} leave it empty if necessary  
   {Probability distribution of densities is a fundamental
     measure of molecular cloud structure, containing information on how the material arranges itself in molecular clouds.}
  % aims heading (mandatory)
   {We derive the probability density functions (PDFs) of column density for a
     complete sample of prominent molecular cloud complexes closer than $d \lesssim 200$ pc. For comparison, additional complexes at $d\approx 250-700$ pc are included in the study.}
  % methods heading (mandatory)
   {We derive near-infrared dust extinction maps for 23 molecular
     cloud complexes, using the \textsf{nicest} colour excess mapping
     technique and data from the 2MASS archive. The extinction maps
     are then used to examine the column density PDFs in the clouds.}
  % results heading (mandatory)
   {The column density PDFs of most molecular clouds are well-fitted by log-normal functions at low column densities ($0.5$ mag $< A_V \lesssim 3-5$
     mag, or $-0.5 < \ln{A_V/\overline{A}_V} \lesssim 1$). However, at higher column
     densities prominent, power-law-like wings are common. In
     particular, we identify a trend among the PDFs: active
     star-forming clouds always have prominent non-log-normal wings. In contrast, clouds without active star formation resemble log-normals over the whole
     observed column density range, or show only low excess of higher column densities. This trend is also reflected in the cumulative
     forms of the PDFs, showing that the fraction of high column density
     material is significantly larger in star-forming clouds. These observations
     are in agreement with an evolutionary trend where turbulent motions are the main
     cloud-shaping mechanism for quiescent clouds, but the density
     enhancements induced by them quickly become dominated by gravity (and
     other mechanisms) which is strongly reflected by the shape of the column
     density PDFs. The dominant role of the turbulence is restricted
       to the very early stages of molecular cloud evolution, comparable to
       the onset of active star formation in the clouds.}
  % conclusions heading (optional), leave it empty if necessary 
    {}
   \maketitle
%
%________________________________________________________________

\section{Introduction}

% The First Paragraph. 
%---------------------

Star formation takes place exclusively in molecular
clouds, or more precisely, in the most extreme density enhancements of those
clouds.
In the current view,
the structure of molecular clouds, and thereby the occurrence of the density
enhancements, is heavily affected by the motions induced by supersonic turbulence \citep[e.g.][]{sca98}. In parallel, the
cloud structure is also crucially affected by the self-gravity of gas and magnetic
fields inside the clouds. The relative strengths of these cloud-shaping mechanisms are currently under lively debate and regarded as one of the
critical open questions in the physics of the interstellar medium \citep[for
reviews, see][]{mac04, mck07}.

% The theoretical prediction.
%----------------------------

The impact of supersonic turbulence for molecular cloud structure is
concretely evidenced by the structural characteristics of molecular
  clouds that seem to agree with
   theoretical predictions and numerical simulations of
   such turbulence \citep[see e.g. \S 2.1 in][]{mck07}. %These characteristics include a wide variety of statistics
%   of both density and velocity \citep[see e.g. \S 2.1 in][]{mck07}.
   One particularly important statistical property is the probability distribution of
   densities, which describes the probability of a volume $dV$ to have a density
   between $[\rho,\rho + d\rho]$. This distribution is expected to take a
   log-normal shape in isothermal, turbulent media not significantly affected by the self-gravity of gas
   \citep[e.g.][]{vaz94, pad97, ost99}. The function plays a fundamental role in
   current theories of star formation: it is used to explain among others
   the initial mass function of stars, and the star formation
   rates and efficiencies of molecular clouds \citep[e.g.][]{pad02, elm08}. 

% Column density PDF
%-------------------

The log-normality of the probability distributions of density is reflected also in \emph{column} densities
computed from simulations \citep[e.g.][]{ost01, vaz01, fed09}. Unfortunately, measuring column densities in molecular clouds is a challenge
in astrophysics in itself. The commonly used methods for deriving column
densities, i.e. measurements of CO line emission, thermal dust emission, and
dust extinction, suffer from various model-dependent effects, and often probe
only narrow ranges of physical conditions \citep[e.g.][]{goo09, vas09}. 

% About extinction mapping
%--------------------------

For the most nearby molecular clouds, dust extinction measurements in the near-infrared provide sensitivity over a relatively
wide dynamical range, starting from $N(\mathrm{H}_2+\mathrm{H}) \gtrsim 0.5 \times
10^{21}$ cm$^{-2}$ \citep{lom06}. The highest measurable
column densities depend on the limiting magnitude
of near-infrared data available; using e.g. 2MASS data, column densities of $\sim 25 \times 10^{21}$  cm$^{-2}$
are
reached \citep[e.g.][]{kai06, lom06}. This broad range, together with the
independency of such data on the dust temperature, makes dust extinction mapping a viable
method to study the large-scale, lower-density regions of molecular clouds and thereby to test the predictions from simulations of supersonic turbulence.

In this Letter, we present the first results of a study where we utilise a novel
near-infrared dust extinction mapping method to study the structural
  parameters in a large sample of nearby molecular clouds. In
  this paper, we focus on the column density PDFs in the clouds, the presentation of the maps
and further analysis is left to a forthcoming paper (Kainulainen et al. in
prep.). Our cloud sample forms a
complete set of prominent cloud complexes within $200$ pc that have an
extent of more than $\sim 4$ pc, or are roughly more massive than $\sim 10^3$
M$_{\sun}$. For comparison, the sample includes some
  additional clouds up to $d\approx 700$ pc. The method we adopt allows us to
determine the column densities over a range that extends to significantly higher column
densities than can be probed by CO line emission (due to the freeze-out of
molecules), enabling study of the structural parameters in a regime not widely
accessed before. 

%__________________________________________________________________

%******************************************
\section{Extinction mapping method}
\label{sec:methods}
%******************************************

   \begin{figure*}
   \centering
   \includegraphics[bb=0 20 550 240, clip=true, width=0.9\textwidth]{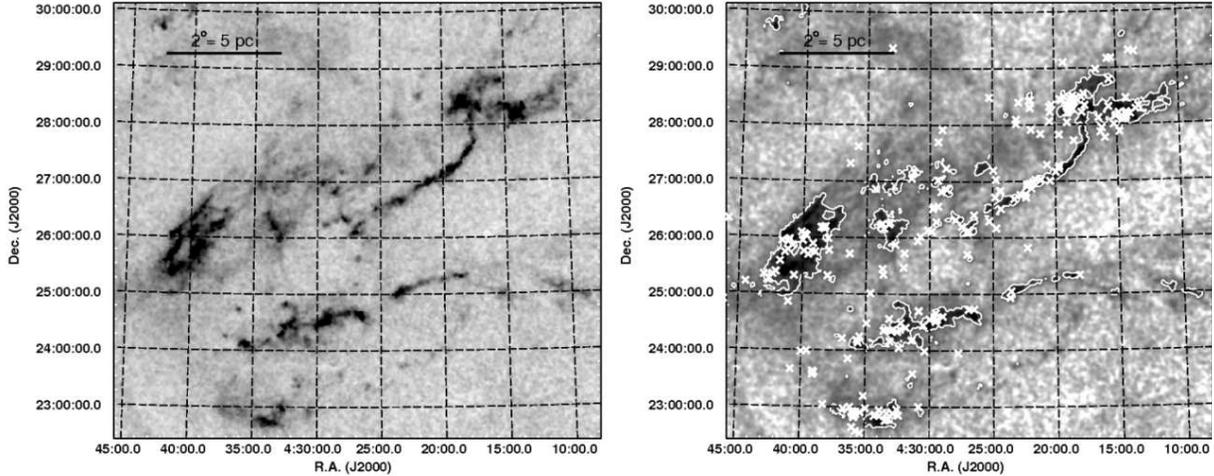}
      \caption{\textbf{Left: }Wide-field extinction map of the Taurus molecular cloud complex covering $\sim 7.5^\circ \times 7.5^\circ$ ($\sim 18 \times 18$
        pc at $d=140$ pc). The FWHM resolution of the map is
        $2.4\arcmin$. \textbf{Right: }The same, but in logarithmic scaling
        highlighting the low column density regions. The contour at $A_V=4$
        mag shows the region above which the column density PDF differs from
        the simple log-normal form. The crosses show the embedded population
        of the cloud as listed by \citet{reb09} (colour figures given in the online version).
              }
         \label{fig:taurus}
   \end{figure*}
   
   \onlfig{1}{
   \begin{figure*}
   \centering
   \includegraphics[bb=0 20 550 240, clip=true, width=0.99\textwidth]{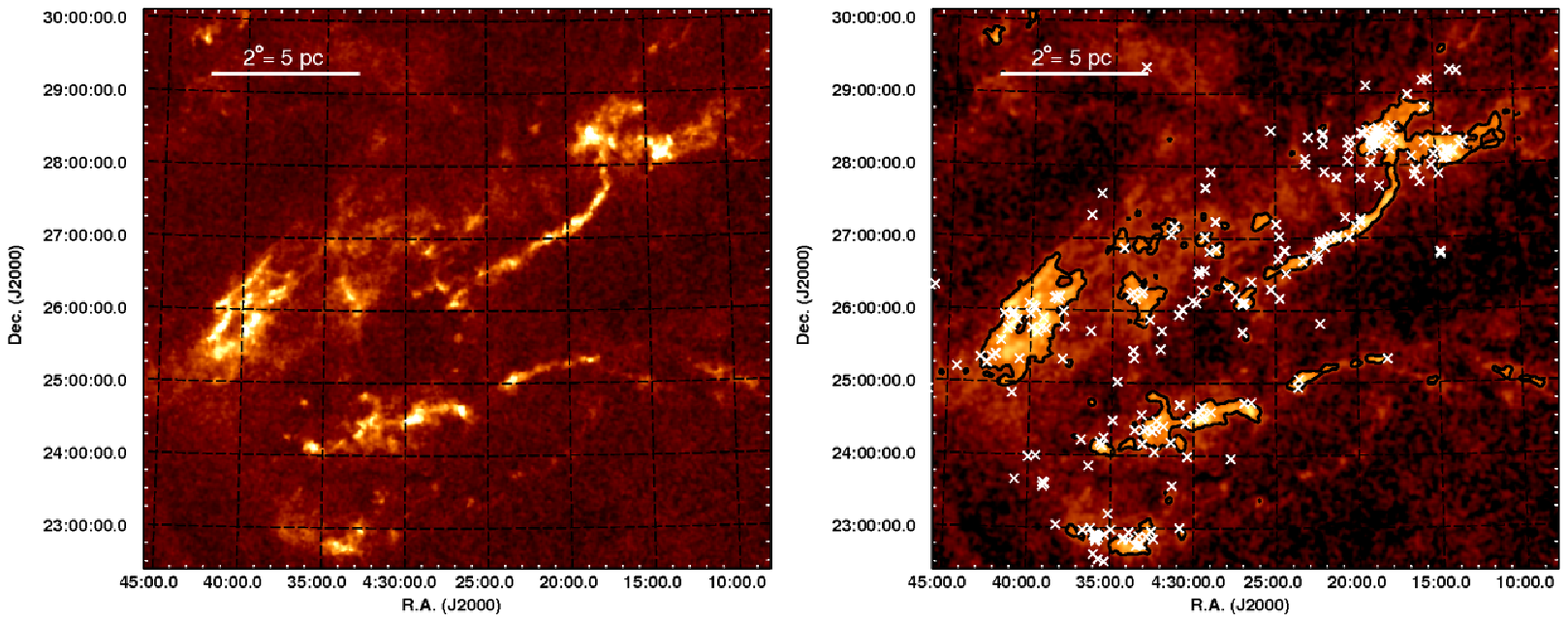}
      \caption{\textbf{Left: }Wide-field extinction map of the Taurus molecular cloud complex covering $\sim 7.5^\circ \times 7.5^\circ$ ($\sim 18 \times 18$
        pc at $d=140$ pc). The FWHM resolution of the map is
        $2.4\arcmin$. \textbf{Right: }The same, but in logarithmic scaling
        highlighting the low column density regions. The contour shows the
        value of $A_V=4$ mag above which the column density PDF differs from
        the simple log-normal form. The crosses show the embedded population
        of the cloud as listed by \citet{reb09} (colour figures given in the online version).
              }
         \label{fig:taurus}
   \end{figure*}}

% Introduce NICEST

We employed the near-infrared dust extinction mapping
technique {\sf nicest} \citep{lom09} to derive
extinction maps of nearby molecular clouds. The method was used
in conjunction with $JHK_S$ band photometric data from 2MASS \citep{skr06}. In
{\sf nicest}, the near-infrared colours of stars, shining through molecular clouds, are compared to
the colours of stars in a nearby reference field that is free from
extinction and in which the brightness distribution of stars is similar to the on-cloud region. This comparison yields estimates of near-infrared extinction
towards the stars in the molecular cloud region. The extinction values are
then used to compute a spatially smoothed map of extinction through the
cloud. In the following, we introduce our practical implementation of the
method. For the further description of the method itself, we refer to \citet{lom01} and \citet{lom09} \citep[see also][]{lom05}. 

% Details of the implementation.

We applied {\sf nicest} to several fields covering previously known molecular
cloud complexes. The clouds included in the study are listed in Table \ref{table:1}. As an example, Fig. \ref{fig:taurus} shows the extinction map of the Taurus complex.
In order to directly compare the maps of different clouds,
their physical resolution was selected to be 0.1 pc ($2\arcmin$ at 170 pc distance). This selection corresponds to the Jeans length for
a core at $T=15$ K and $\overline{n}=5 \times 10^4$ cm$^{-3}$. The distances adopted for the clouds are listed in
Table \ref{table:1}. For the most clouds farther away than
  200 pc, we used a physical resolution of $0.6$ pc. The PDFs of
these clouds are not directly comparable to those whose resolution is 0.1 pc.

Stars that are either embedded inside the cloud or on the foreground with
respect to it can bias the derived extinction. To minimize the
contribution of such sources, we used catalogues of
previously identified cloud members from the literature to
directly remove sources. In addition, we used the ``sigmaclipping'' iteration, i.e. each source towards which the estimated extinction
differed by more than 5-sigma from the local average was removed from the
sample. Another possible source of bias in the data is that the background stellar density
varies among the clouds, according to their galactic coordinates. We
investigated the possible effect of this on the PDFs by degrading the background stellar density of some clouds that
are close to the galactic plane and recomputing the extinction maps. As these
experiments had no impact on the results shown in Sect. \ref{sec:results}, we
did not include any correction for the differences.

% About the accuracy of the method (this is an important point here!!)

The noise in the extinction maps depends on the galactic
coordinates and on extinction. Typically, the
$3\sigma$-error at low column densities is $0.5-1.5$ mag. The extinction measurements ``saturate'' at about $A_V = 25$ mag. We note that the fractional area
where $A_V \gtrsim 25$ mag is small and we are not likely to significantly miss
mass due to an inability to probe higher extinctions.

   \begin{figure*}
   \centering
   \includegraphics[bb=0 7 350 160, clip=true, width=0.43\textwidth]{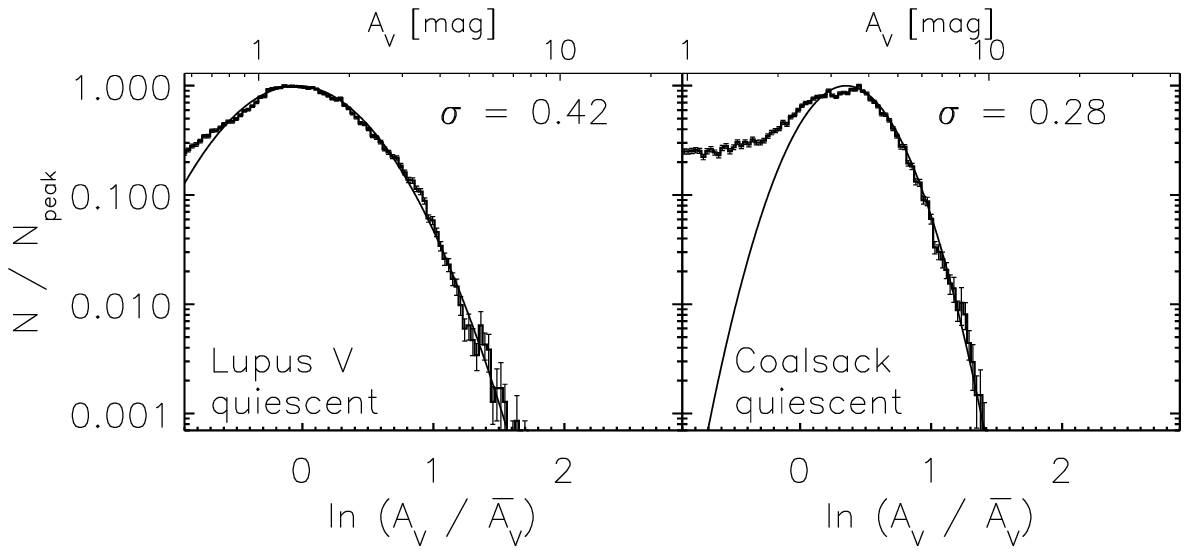}
   \includegraphics[bb=0 7 350 160, clip=true, width=0.43\textwidth]{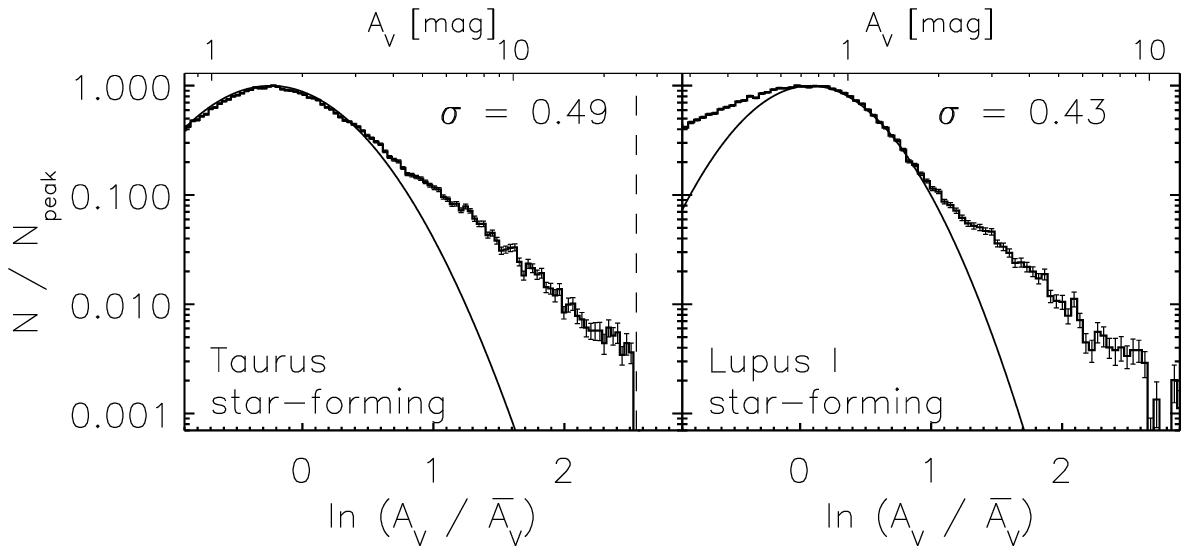}
      \caption{\textbf{Left: } Probability density
        functions (PDFs) of the column density for the non-star-forming clouds Lupus 5 and
        Coalsack. \textbf{Right: } The same for the active star-forming clouds Taurus and Lupus 1. The error bars show the $\sqrt{N}$
        uncertainties. Solid lines show the fits of log-normal functions to the distributions around the peak,
        typically over the range $\ln{A_v/\overline{A}_V}=[-0.5, 1]$. The
        dispersions of the fitted functions are shown in the panels. The
        x-axis on top of the panels shows the extinction scale in
        magnitudes. The vertical dashed line shows the upper limit of
        extinction values probed by the extinction mapping method. Similar plots for 19 other clouds
        are shown in Figs. \ref{fig:pdfs2}-\ref{fig:pdfs4} (online only).    }
         \label{fig:pdfs}
   \end{figure*}

%
%______________________________________________________________

%************************
\section{The column density PDFs for nearby clouds}
%************************
\label{sec:results}

Figure \ref{fig:pdfs} shows the mean-normalised PDFs of logarithmic column
densities for four clouds of the study. Figures
\ref{fig:pdfs2}-\ref{fig:pdfs4} show the PDFs for 19 other clouds (online only). 
In these figures and throughout this Letter, we have divided our sample in
active and non-active star-forming clouds based on the presence of confirmed 
young stellar objects in the clouds.

Qualitatively, most PDFs show a log-normal-like peak, followed by a power-law-like
extension at higher column densities. The strength of the
extension varies, being dominant for some clouds
(e.g. Taurus) and absent for others (e.g. Coalsack). For some
  clouds, the PDF differs from a log-normal shape also at very low column
  densities (see Sect. \ref{sec:discussion}). Directed by theoretical predictions, we fitted the peaks
  of the PDFs using log-normal functions:
\begin{equation} 
p(s) \sim \exp\big[-\frac{(\ln{A_V / \overline{A}_V} - m)^2}{2\sigma^2}\big], 
\label{eq:lognormal}
\end{equation}
where $\overline{A}_V$ is the mean extinction, and $m$ and
  $\sigma$ are the scale and dispersion in logarithmic units. The fits are shown in Figs. \ref{fig:pdfs} and \ref{fig:pdfs2}-\ref{fig:pdfs4}. Since it is evident that most PDFs
are not well fitted by log-normals over their entire range,
the fit was typically made over the range $s=[-0.5, 1]$.  
The dispersions of the fitted log-normal functions are shown in Table \ref{table:1}, and they span the
range $\sigma_s \approx 0.3-0.5$. Table \ref{table:1} also shows the total
mass , and the mean and standard deviation of the pixels above $A_V=0$ mag. The total mass was calculated by summing up the extinction values in the map above
$A_V=0.5$ mag and adopting the standard ratio of $N(\mathrm{H}_2+\mathrm{H})/A_V = 9.4 \times 10^{20}$ cm$^{-2}$ \citep{boh78}.

% Cumulative mass fraction

\begin{figure}
   \centering
   \includegraphics[bb=41 11 410 313, clip=true, width=0.65\columnwidth]{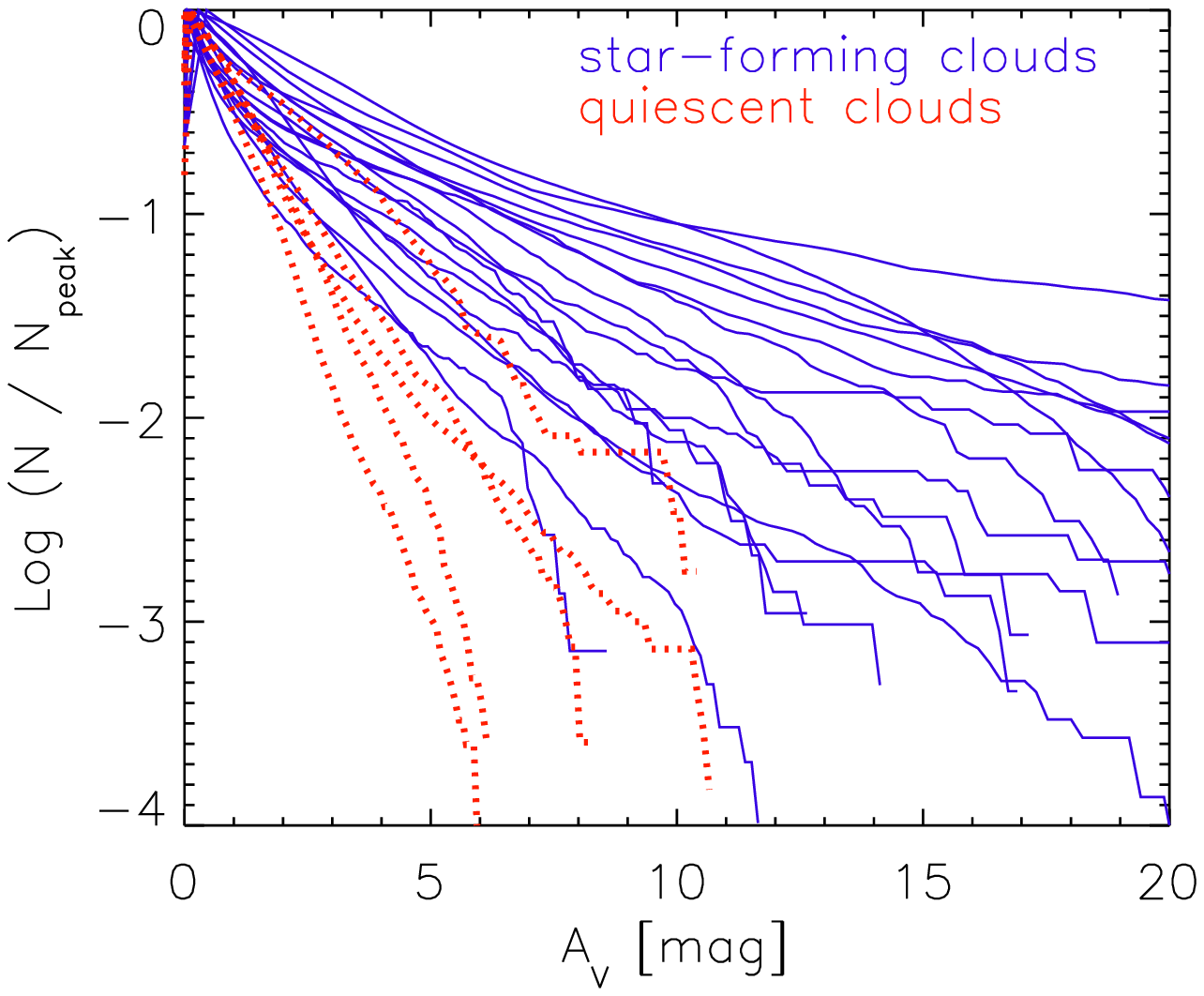}
      \caption{Cumulative forms of the PDFs shown in Figs. \ref{fig:pdfs} and \ref{fig:pdfs2}-\ref{fig:pdfs4}. The
        curves show the fractional mass above the certain extinction threshold
        (abscissae). Solid blue lines are for clouds that show active star formation
        and dotted red lines for clouds without active star formation.
              }
         \label{fig:cmf-comb}
   \end{figure}

Another interesting form of the PDFs is the cumulative form of the pixel
probability distribution, describing the fractional mass enclosed by an
isocontour as a function of column density (or more precisely, the
  \emph{survival function}). The cumulative PDFs are shown in Fig. \ref{fig:cmf-comb} for all the clouds of the study. 
In this figure, the active star-forming clouds are separated from quiescent
clouds. Clearly, the fraction of mass in high column density regions is higher
in star-forming clouds than in clouds without star formation. We approximated the
average cumulative functions for these two classes as a simple mean of all the
clouds in the class, which resulted in the relation $(N /
N_\mathrm{peak})_{\mathrm{SF}} \approx (N / N_{\mathrm{peak}})_{\mathrm{non-SF}}^{0.4}$. For example, the star-forming
clouds then have roughly one order of magnitude more mass above $A_V=5$ mag
than non-star-forming clouds and more than three orders of magnitude above
$A_V=15$ mag.

% Figure of the CMFs combined

%********************************************
\section{Discussion and conclusions}
\label{sec:discussion}
%********************************************

% First paragraph
%----------------

While supersonic turbulence is expected to develop a density PDF close to a
log-normal distribution, prominent deviations from that shape are predicted in
strongly self-gravitating systems \citep[e.g.][]{kle00,fed08}. Recent
observational studies have indeed indicated that the column density PDFs of
molecular clouds are
close to log-normal distributions. For example, \citet{lom06, lom08} examined the
column density PDFs in the Pipe, Rho Oph,
and Lupus molecular clouds. They concluded that the PDF of Ophiuchus is
satisfactorily fitted by a log-normal function, while the PDF of Lupus is extremely well
fitted by it. However, the PDF of Pipe required four log-normal
functions, which they suggested originated from physically distinct
components along the line of sight. 

\begin{table}
\begin{minipage}[t]{\columnwidth}
\caption{Molecular cloud properties and the derived parameters}
\centering
\renewcommand{\footnoterule}{}  % to avoid a line before footnotes
\begin{tabular}{lcccccc}
\hline \hline
Cloud & $D$\footnote{References: (1) \citet{lom08b} (2) \citet{tor07} (3)
\citet{str96} (4) \citet{knu98} (5) \citet{cas98} (6) \citet{kun98} (7)
\citet{lom06} (8) \citet{mat79} (9) \citet{oba98} (10) \citet{cor97} (11)
\citet{bur93} (12) \citet{cer93} (13) \citet{cra70} (14) \citet{lad09}.} &
$M_{H2} \times 10^4$ & $\overline{A_V}$ & $\sigma_{data}$ & $\sigma_{fit}$ &
$N_\mathrm{YSO}$\footnote{An order-of-magnitude estimate of the pre-main-sequence
  star population in the clouds, based on \citet{rei08}.} \\ 
\hline
\multicolumn{6}{c}{Star-forming clouds, physical resolution 0.1 pc}\\
\hline                      
\object{Ophiuchus}        &  119$^1$ & 0.6  & 2   & 4.3 & 0.48\footnote{No
  good fit was achieved. A rough estimate is given for reference.} &  300 \\
\object{Taurus}           &  140$^2$ & 0.81 & 1.8 & 1.5 & 0.49 &  300 \\
\object{Serpens}          &  259$^{3}$ & 3.6  & 3 & 2.4 & 0.51 &    300 \\
\object{Cha I}            &  165$^4$ & 0.23 & 1.3 & 1.3 & 0.35 & 200 \\
\object{Cha II}           &  150$^4$ & 0.15 & 1.3 & 1.0 & 0.35 & 50  \\
\object{Lupus III}        &  155$^1$ & 0.11 & 1.3 & 0.9 & 0.35$^c$ & 50 \\
\object{CrA cloud}        &  129$^5$ & 1.2  & 0.8 & 1.0 & 0.44 & tens \\
\object{Lupus I}          &  155$^1$ & 0.29 & 1.0 & 0.7 & 0.43 & tens \\
\object{LDN1228}$^d$      &  200$^{6}$ & 0.23 &  1.1 & 0.7 &  0.32 &  tens  \\
\object{Pipe}             &  130$^7$ & 0.95 & 2.5 & 1.4 & 0.29$^c$ & 15 \\
\object{LDN134}           &  100$^{8}$ & 0.13 & 1.2 & 0.6 & 0.39 &  a few \\
\object{LDN204}\footnote{The most prominent Lynds dark nebula in the region.}
&  119$^1$ & 0.43 & 1.9 & 0.8 & 0.41 &  a few  \\   
\object{LDN1333}$^d$      &  180$^9$ & 0.57 & 1.2 & 0.5 & 0.38 &  a few    \\
\hline
\multicolumn{6}{c}{Non-star-forming clouds, physical resolution 0.1 pc}\\
\hline                      
%LDN1358          &  258, cf. Kun 2008 & 0.64 & 1.7 & 0.9 & 0.38 \\
\object{LDN1719}$^d$      &  120$^4$ & 0.53 & 0.6 & 0.7 & 0.50 &   \\
\object{Musca}            &  150$^4$ & 0.07 & 1.0 & 0.7 & 0.45 &   \\
\object{Cha III}          &  150$^4$ & 0.18 & 1.3 & 0.8 & 0.46 &   \\ 
%LDN673           &  300     &   &  &  &  \\
\object{Coalsack}         &  150$^{10}$     & 0.5  & 2.7 & 1.4 & 0.28 &   \\
\object{Lupus V}          &  155$^1$ & 0.36 & 1.4 & 0.7 & 0.42 &   \\
\hline 
\multicolumn{6}{c}{Star-forming clouds, physical resolution 0.6 pc}\\
\hline                      
\object{Ori A GMC}        &  450$^{11}$ & 11   & 1.4 & 2.8 & 0.5 &  $>$ 2000 \\
\object{Per cloud}        &  260$^{12}$ & 2.0  & 1.7 & 1.7 & 0.49 &  $>$ 100 \\
\object{Ori B GMC}        &  450$^{11}$ & 9.0  & 1.2 & 1.7 & 0.59 &  $>$ 100 \\
\object{Cepheus A}        &  730$^{13}$ & 3.5  & 1.5 & 1.6 & 0.47 &  $>$ 100 \\
\object{California}       &  450$^{14}$ & 11   & 1.4 & 0.7 & 0.51 &  tens\\
\end{tabular}
\label{table:1}
\end{minipage}
\end{table}

% Discussion on PDFs
%-------------------

The column density PDFs presented in this Letter show that simple log-normal
functions fit the PDFs poorly when considering the whole observed column
density range, i.e. $N=0.5- 25 \times 10^{21}$ cm$^{-2}$.  
As seen in Figs. \ref{fig:pdfs} and \ref{fig:pdfs2}-\ref{fig:pdfs4}, the PDFs
of most clouds deviate from simple log-normality at
$s \gtrsim 0.5-1$, or $A_V \gtrsim 2-5$ mag. Even though
the log-normals fit the peaks of the PDFs well, the
excess ``wings'' at higher column densities are persistent features of the
molecular cloud structure. Likewise, about half of the clouds show
  non-log-normal features at low column densities. It is, however, difficult
  to ascertain whether the low column density features are real. It is quite
  possible that they are mostly residuals caused by other, physically distinct
  clouds along the line of sight. Nevertheless, they can also be real signatures of
  cloud structure; non-log-normal features at low column densities have been
  predicted, related to intermittent fluctuations in turbulent media \citep[e.g.][]{fed09}. %This result seems
%to be somewhat in contradiction to the Lombardi et al. studies. However, it is difficult to
% ascertain whether there is an overabundance of higher column
%   densities in the PDFs
% presented by Lombardi et al., since it is not detectable, if present,
%   when the PDF is plotted with linear axis scaling \citep[see Figs. 13 and
% 14 in][]{lom08}. 
   We note that the number of pixels in the non-log-normal, higher extinction parts is not small. In fact, the threshold above
   which the wings become prominent ($A_V \gtrsim 2-5$ mag) is rather low,
   suggesting that material related to star formation is dominantly located in the non-log-normal part of the PDF. This is illustrated in the right panel of 
   Fig. \ref{fig:taurus} which shows the extinction map of Taurus with a
   contour at $A_V=4$ mag highlighting the regions belonging to the
   non-log-normal wing of the PDF. The figure also shows the known
   pre-main-sequence stars that clearly concentrate on the regions of high column density \citep{reb09}.

%\textbf{The origin of the wings at very low column densities is not clear, but
%can be speculated with. It is quite likely that some
%regions have contamination from other clouds along the line of sight. Such
%clouds, especially at larger distances, would increase particularly the
%relative amount of low column densities. This seems a likely explanation for
%clouds close to the Galactic plane. It is, however, possible that the features
%are real signatures of cloud structure. In fact, non-log-normal features at
%low column densities have been detected in simulations of supersonic
%turbulence, and they have been related to the intermittent fluctuations in the
%turbulent media \citep[e.g.][]{fed09}.}

Intriguingly, our data appear to show a clear trend. All clouds with active star formation show
strong excess of higher column densities (Figs. \ref{fig:pdfs} and \ref{fig:pdfs2}). In
contrast, almost all quiescent clouds have PDFs that either are well
described by log-normal functions
  over the entire column density range, or show relatively low excess of
high column densities (Figs. \ref{fig:pdfs} and \ref{fig:pdfs3}). The
  trend is obvious for the lower-mass clouds in our sample,
  but an indication of
it is seen also among the more massive clouds (Fig. \ref{fig:pdfs4}): the active star-forming clouds
of Orion have prominent wings compared to the California nebula, a massive
cloud with significantly lower star-forming activity \citep{lad09}.
In the context of turbulent molecular cloud evolution, these observations are in agreement with a
picture in which the structure of a molecular cloud
in the early stage of its evolution is decisively shaped by turbulent
motions. Hence, its column density PDF is
  close to log-normal, like it is the
case for the non-active clouds in our sample. As the cloud evolves,
  prominent local density enhancements can become self-gravitating, which
  also assembles a growing fraction of the gas to higher column density structures. This significantly alters the simple
log-normal form of the column density PDF, and is very concretely demonstrated
by the cumulative forms of the PDFs (Fig. \ref{fig:cmf-comb}), showing how dramatically the fraction of
material at high column density increases from non-active to active
clouds. The cumulative PDFs shown in Fig. \ref{fig:cmf-comb} generalize this
trend, suggested earlier by studies of individual cloud complexes \citep{cam99,
  lom06, lom08, lad09}. In the low column density regions of molecular
clouds turbulent motions prevail as the dominant structure-shaping mechanism,
as indicated by the log-normal-like parts of the PDFs. This
appears natural, since those are likely to be the regions where the role of self-gravity remains small.

% Concluding paragraph
%---------------------

In this Letter, we have characterised the shape of the
column density PDFs in nearby molecular clouds and
demonstrated the prevalence of non-log-normalities in them. From the PDFs, we
identified a trend that is in agreement with a picture where self-gravity has a
significant role in shaping the cloud structure starting from a very early
stage, corresponding to the formation of first stars in the
cloud. An immediate question following these observations is to what
  extent similar features are present in the simulations of supersonic turbulence. This can be directly
addressed by a comparison of our data to simulations that include self-gravity
and follow the evolution of cloud structure as a function of time \citep[e.g.][]{off08, ban09}. 
The data presented in this Letter provide a unique set for
  this purpose, and we are going to address this in a forthcoming paper.

%______________________________________________________________

\begin{acknowledgements}

The authors would like to thank the referee, L. Cambr{\'e}sy, for the comments
that improved the paper, and C. Federrath, R. Banerjee, M.-M. Mac Low, and
R. Klessen for hepful discussions. This research has made use of the NASA/IPAC Infrared Science Archive,
which is operated by the Jet Propulsion Laboratory, California Institute of
Technology, under contract with the National Aeronautics and Space
Administration. This publication makes use of data products from the Two
Micron All Sky Survey, which is a joint project of the University of
Massachusetts and the Infrared Processing and Analysis Center/California
Institute of Technology, funded by the National Aeronautics and Space
Administration and the National Science Foundation. 
\end{acknowledgements}

%-------------------------------------------------------------

%\section{PDFs of 22 molecular clouds}

   \onlfig{4}{
   \begin{figure*}
   \centering
   \includegraphics[width=\textwidth]{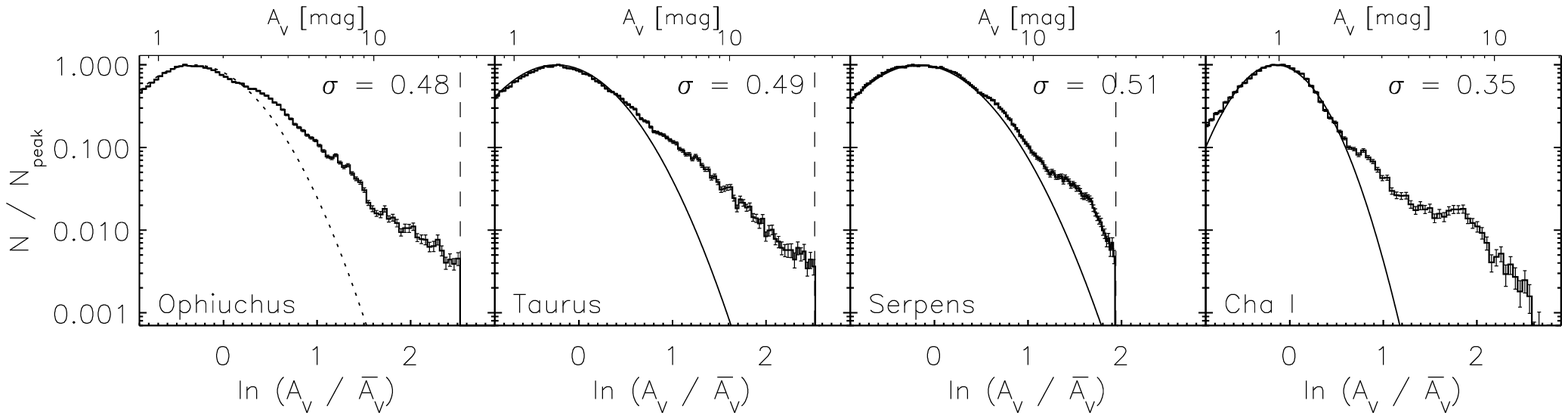}
   \includegraphics[width=\textwidth]{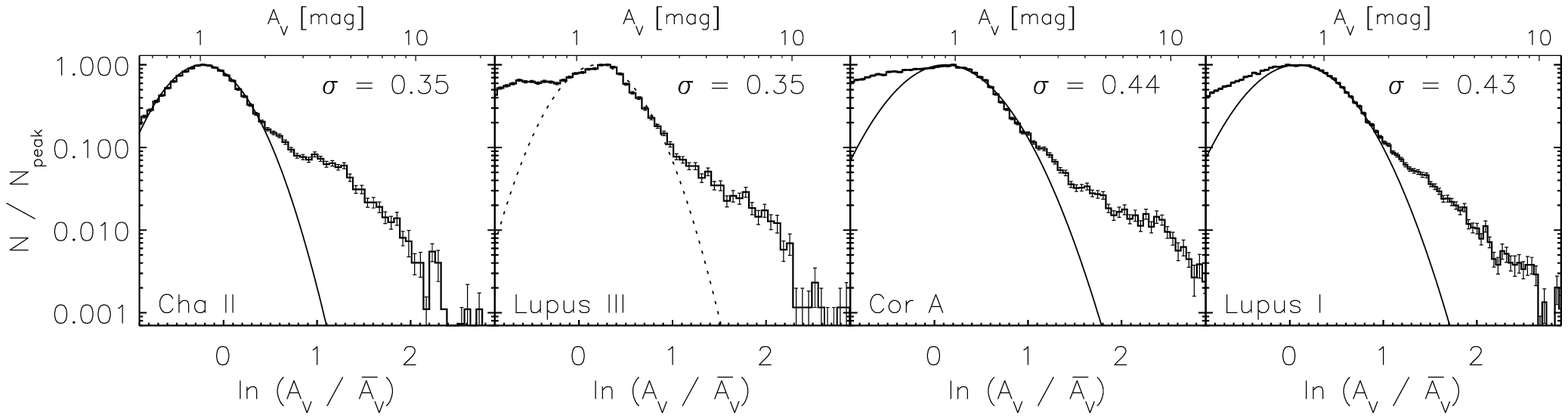}
   \includegraphics[width=\textwidth]{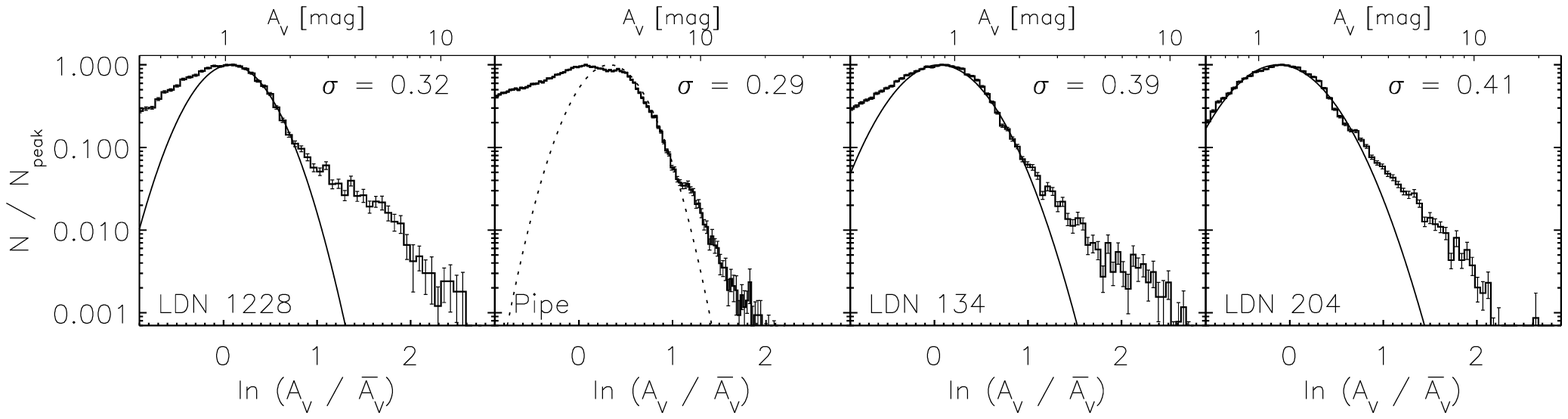}
   \includegraphics[width=\textwidth]{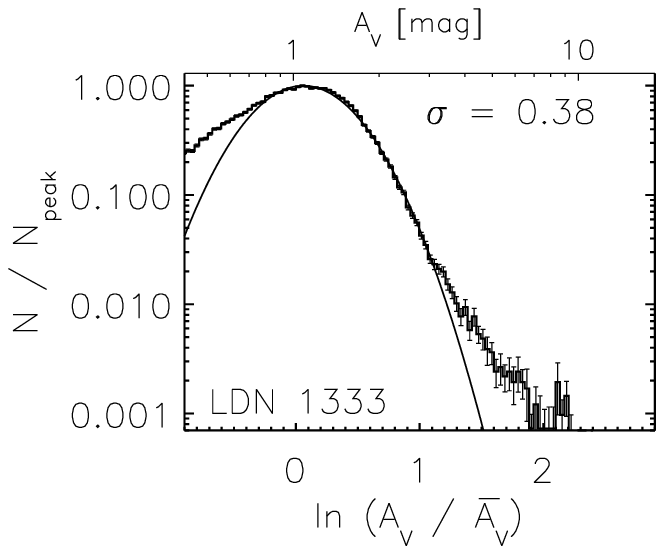}
      \caption{Probability density functions (PDFs) of a normalised column
        density for 13 star-forming clouds in the study. The error bars show the $\sqrt{N}$
        uncertainties. Solid lines show the fits of lognormal functions to the distributions around the peak,
        typically over the range \textbf{$\ln{A_v/\overline{A}_V}=[-0.5,1]$}. The
        dispersion of the fitted function is shown in the panels. For
        some clouds, no reasonable fit was achieved over any $A_V$ range. For
        those clouds, we show for a reference a function approximating the shape using a dotted line. The
        x-axis on top of the panels shows the extinction scale in magnitudes.  
              }
         \label{fig:pdfs2}
   \end{figure*}}

  \onlfig{5}{
  \begin{figure*}
   \centering
   \includegraphics[width=\textwidth]{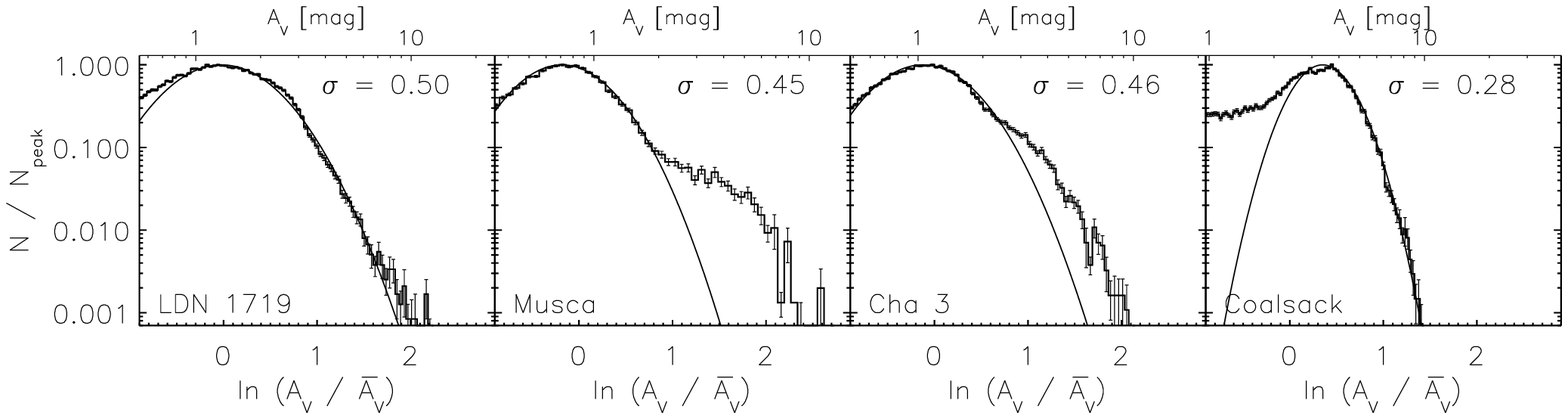}
   \includegraphics[width=\textwidth]{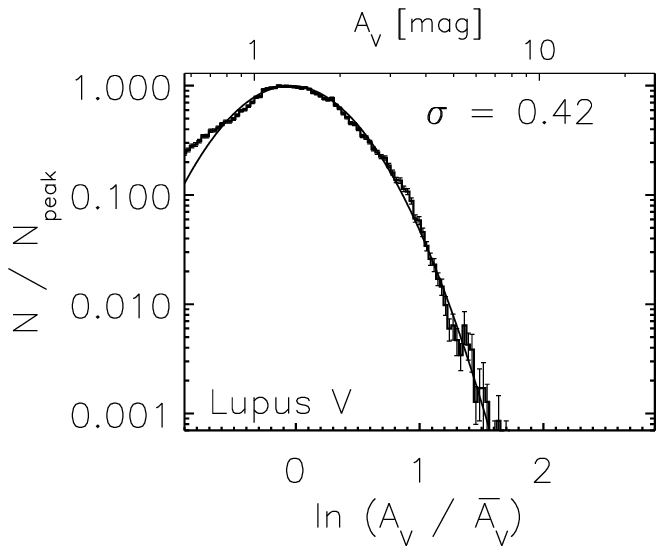}
      \caption{Same as Fig. \ref{fig:pdfs2}, but for clouds we classify as
        clouds not containing active star formation.}
         \label{fig:pdfs3}
   \end{figure*}}

   \onlfig{6}{
   \begin{figure*}
   \centering
   \includegraphics[width=\textwidth]{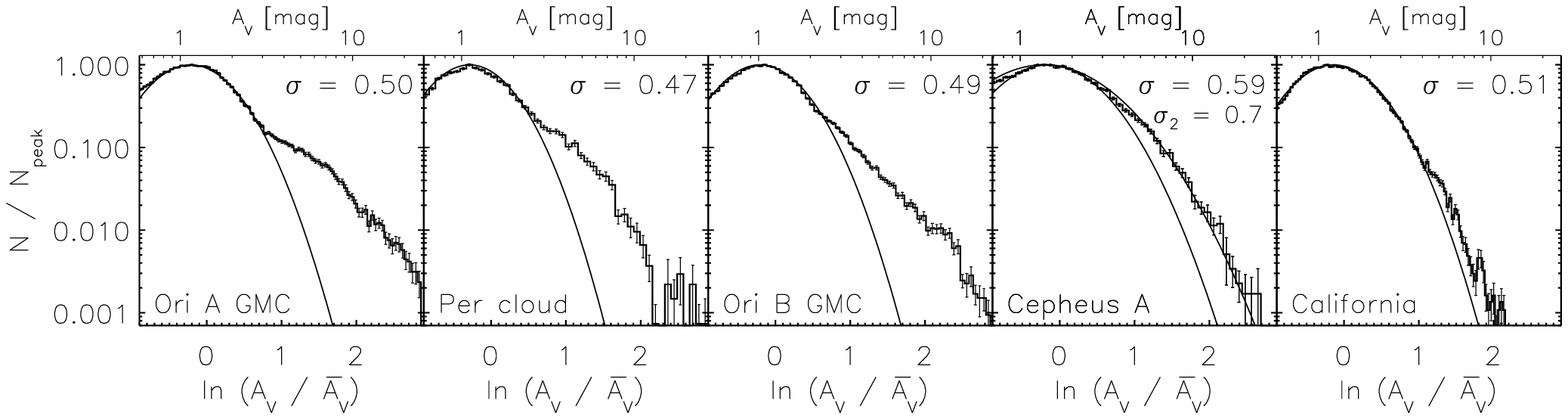}
      \caption{Same as Fig. \ref{fig:pdfs2}, but for clouds at varying
        distances of 250-700 pc. The PDFs are smoothed to the common physical
        resolution of 0.6 pc. For Cepheus, two equally good fits are shown.
              }
         \label{fig:pdfs4}
   \end{figure*}}

\end{document}